# 2.5-D ELECTRICAL RESISTIVITY FORWARD MODELLING WITH UNDULATING TOPOGRAPHY USING A MODIFIED HALF-SPACE ANALYTICAL SOLUTION

Naveen K[1*], Michael C Koch[2], Kazunori Fujisawa[3], Arindam Dey[1*], Sreedeep S[1]

[1]*Civil Engineering Department, Indian Institute of Technology Guwahati, Guwahati, Assam, India.*

[2]*Research School of Earth Sciences, Australian National University, Canberra ACT 2601, Australia.*

[3]*Graduate School of Agriculture, Kyoto University, Kitashirakawa-Oiwakecho, Sakyo-ku, Kyoto, Japan.*

**SUMMARY**

Field measurements for direct current (DC) resistivity imaging, used for subsurface profiling, are frequently conducted over undulating terrain. Accurately incorporating such topographic variations in its forward modelling is essential for reliable inversion and interpretation. Singularity removal techniques provide a computationally efficient framework by analytically representing the singular component of the electric potential. Existing secondary potential formulations use the analytical solution for a flat homogeneous half space, but this assumption is realistic only when the source lies on a locally smooth, flat planar surface. In practice, natural topography often contains sharp corners or regions of high curvature, and additional slope discontinuities arise from linear finite element discretization. These conditions invalidate the flat-surface analytical primary field and lead to substantial modelling errors. These errors originate from a fundamental geometric mismatch between the flat half-space analytical primary field and the true solid angle subtended by the topography at the source. This study presents an improved singularity removal strategy for 2.5-D forward modelling by deriving a new analytical primary potential for a V-shaped wedge. The formulation remains valid for sharply varying surfaces and accurately captures the singular behaviour without requiring geometric smoothing or excessive mesh refinement. By embedding the correct geometric singularity into the primary field, the proposed formulation remains consistent with both the discretized surface geometry and the physical boundary conditions. Numerical experiments on flat, V-shaped trench, and sinusoidal hill-valley models reveal that the proposed method consistently achieves errors below 0.1 per cent, even when using coarse linear finite element meshes. The approach provides a robust and computationally efficient alternative for DC resistivity modelling in complex undulating terrain.

**Key words**: Electrical resistivity tomography (ERT); Electromagnetic theory; Finite element method; Numerical modelling; Numerical solutions

# 1    INTRODUCTION

Direct current (DC) resistivity imaging is a prominent geophysical method employed for near-surface investigation in archaeological, geophysical, hydrogeological, environmental and geotechnical fields (M.H. Loke et al. 2013). Typical applications include detecting cavities, structural anomalies, contaminant plumes, groundwater features, and seepage within embankments and dams. In practice, measurement profiles are often acquired over undulating terrain, making accurate representation of surface topography essential for reliable inversion and interpretation. The electric potential field is sensitive to geometric distortions of the ground surface, and numerical inaccuracies in its computation directly degrade the recovered conductivity distribution. Several studies have shown that neglecting or oversimplifying topography leads to substantial artefacts in the inverted models, including displaced or smeared anomalies, incorrect resistivity contrasts, and misinterpretation of subsurface structures (R.C. Fox et al. 1980; H.T. Holcombe & G.R. Jiracek 1984; P.I. Tsourlos, J.E. Szymanski & G.N. Tsokas 1999; T. Günther, C. Rücker & K. Spitzer 2006; I. Demirci, E. Erdogan, & M.E. Candansayar 2012; L.W. Qiu et al. 2024a). Consequently, inclusion of actual undulating topography in the forward modeling is essential.

The numerical forward modelling of the electric field in DC resistivity imaging has evolved through three principal approaches, namely the integral equation method (IEM), the finite difference method (FDM), and the finite element method (FEM). Integral equation and boundary element formulations, introduced by K. Dieter, N.R. Paterson & F.S. Grant (1969) and M. Okabe (1981), are efficient for topologically simple anomalies but are not well suited for arbitrary topography or complex internal geometries. More recently, Z. Ren, H. Chen & J. Tang (2018a) employed a volume-surface integral formulation to incorporate topographic effects into DC resistivity forward modelling. Finite difference schemes, first applied to 2-D and 3-D DC resistivity by I.R. Mufti (1976) and A. Dey & H.F. Morrisson (1979) and later refined by K. Spitzer (1995), require structured orthogonal grids, which cannot accurately represent irregular terrain. Although S. Penz et al. (2013) used a generalised FDM with unstructured meshes, such formulations remain less flexible than FEM for large, complex domains. The finite element method, introduced by J.H. Coggon (1971) and later utilized by several researchers (D.F. Pridmore et al. 1981; B. Zhou & S.A. Greenhalgh 2001; C. Rücker, T. Günther & K. Spitzer 2006; Z. Ren & J. Tang 2014; L.W. Qiu, J.T. Tang & Z.G. Liu 2024b), offers significantly greater adaptability through unstructured meshes that capture arbitrary geometries and allow efficient local refinement near electrodes and conductivity contrasts. Early FEM implementations used block oriented structured meshes, which could accommodate simple topography but lacked refinement efficiency. The adoption of unstructured meshes established FEM as the preferred approach for DC resistivity forward modelling in settings with complex, undulating terrain.

Solving the Poisson type governing equation is challenging due to the singularity in potential at point source electrodes. For flat surfaces, singularity removal techniques introduced by T. Lowry, M.B. Allen & P.N. Shive (1989) and refined by S. Zhao & M.J. Yedlin (1996) split the total potential into a primary potential, corresponding to a point source in a flat homogeneous half-space, and a secondary potential representing conductivity variations. The primary term captures the singular field explicitly through an analytical solution, while the secondary term is computed numerically. For undulating topography, however, no analytical solution for the primary potential exists. Previous works have addressed this by numerically evaluating the primary potential for undulating terrain. C. Rücker et al. (2006) used FEM with second order elements and dense mesh refinement around the source, whereas M. Blome, H. Maurer & S. Kersten (2009) employed a boundary element approach. Although accurate, these methods are computationally expensive, even if the primary field is computed only once per inversion. When the flat half-space assumption is valid, the primary–secondary split delivers substantial savings because the primary field is analytic and a dense mesh is unnecessary; T. Günther et al. (2006) report that computing the primary potential can consume roughly 38 per cent of total inversion time when a dense higher-order mesh is used.

S. Penz et al. (2013) proposed a modified secondary potential strategy in which a reformulated surface boundary condition permits the use of the flat-surface analytical primary potential. This approach has been applied to problems with isotropic and anisotropic conductivity profiles (Z. Ren et al. 2018b; L.W. Qiu et al. 2024b). However, it performs reliably only when the topography near the source is smooth or nearly planar (S. Penz et al. 2013; Z. Ren & J. Tang 2014; Z. Ren et al. 2018b). At slope discontinuities or at locations of high surface curvature the method produces significant errors. To mitigate this, S. Penz et al. (2013) suggested smoothing the topography in the vicinity of the source. This requires modifying the surface geometry, which must be performed manually at each source location, making it laborious and results in deviation from the actual terrain. Alternatively, extremely fine mesh refinement can reduce curvature errors, but this negates the computational advantages of the primary-secondary split and substantially increases the cost of computing the primary potential.

To overcome these limitations, we extend the S. Penz et al. (2013) approach by modifying the analytical expression used for computing the primary potential. A new analytical formulation is derived for cases where the topography near the source exhibits abrupt slope changes. The derivation incorporates the solid angle subtended by the local topographic surface at the source. Replacing the flat-surface solid angle of $2\pi$ with the true geometric solid angle yields a primary potential that accurately represents singular behaviour, even for sharply varying terrain. This modification enables the computation of accurate solutions using linear elements without requiring excessive mesh refinement or geometric smoothing. The effectiveness and performance of the proposed formulation is demonstrated through a series of synthetic tests.

In the remainder of this paper, we first present the governing equation for DC resistivity and its Fourier domain representation for 2.5-D modelling, followed by a review of existing singularity removal techniques for flat and non-flat topography. We then analyse how local curvature influences modelling accuracy under the flat half-space assumption and motivate the need for an improved primary potential formulation. The proposed solid angle based analytical expression for the primary potential is subsequently derived and implemented within a linear FEM framework. Finally, numerical experiments for flat and topographically variable models, including multiple source positions, are presented to validate the accuracy and efficiency of the proposed method and to demonstrate its robustness compared with existing topography handling approaches.

## 2 FORWARD MODELLING OF DC RESISTIVITY

The potential, $u(\mathbf{x})$ at the location $\mathbf{x} = (x, y, z)$, in the domain, $\Omega \subset \mathbb{R}^3$, generated due to a singular current, $I$, injected into the ground through a source electrode at position $\mathbf{x}_s = (x_s, y_s, z_s)$, and with the conductivity distribution of the ground being $\sigma(\mathbf{x})$, is governed by the Poisson's equation

$$\nabla \cdot \mathbf{J}(x, y, z) = \nabla \cdot \big(\sigma(x, y, z)\nabla u(x, y, z)\big) = -I\delta(\mathbf{x} - \mathbf{x}_s) \quad \text{in} \quad \Omega, \tag{1}$$

where $\mathbf{J}$ is the current density. On the surface boundary, $\Gamma_s$, of the domain, $\Omega$, there is no current flow into the air due to its high resistivity and hence a Neumann boundary condition (BC) is applied to this boundary:

$$\mathbf{J} \cdot \mathbf{n} = \sigma \frac{\partial u}{\partial n} = 0 \quad \text{on } \Gamma_s. \tag{2}$$

On the truncated distant boundary, $\Gamma_\infty$, of the subsurface, the mixed boundary condition proposed by A. Dey & H.F. Morrisson (1979) is applied.

$$\sigma\left(\frac{\partial u}{\partial n} + \frac{\cos\theta}{r}u\right) = 0 \quad \text{on } \Gamma_\infty \tag{3}$$

where, $\theta$ is the angle between the outward normal to the boundary and the radial vector from the source to that point, and $r$ is the distance from the source. Assuming the conductivity to be constant in the $y$-direction, and applying a Fourier cosine transform to the potential in the $y$-direction, we obtain an equation in terms of transformed potential, $\tilde{u}(x, k_y, z)$ as

$$\widetilde{\nabla} \cdot \left(\sigma(x,z)\widetilde{\nabla}\widetilde{u}(x,k_y,z)\right) - k_y^2\sigma(x,z)\widetilde{\nabla}\widetilde{u}(x,k_y,z) = -\frac{I}{2}\delta(x-x_s)\delta(z-z_s) \quad (4)$$

where, $\widetilde{\nabla} = \left(\frac{\partial}{\partial x}, \frac{\partial}{\partial z}\right)$ and $k_y$ is the wavenumber related to the $y$ direction. The wavenumbers and corresponding weights are chosen based on the approach discussed in A. Kemna (2000). As per the approach, numerical integration of the inverse Fourier integral is based on a combination of Gauss-Legendre quadrature and Gauss-Laguerre quadrature. The 3-D potential is then obtained by an inverse Fourier transform.

*Singularity removal and boundary condition for non-flat topography*

Since the total potential, $u$, is singular at the source point, numerical approximations typically give poor results in the vicinity of the source location (K.J. Pan & J.T. Tang 2014). The singularity removal technique, introduced by T. Lowry et al. (1989), splits the total potential into a primary potential, $u_p$, resulting from the source current in a uniform half-space with a constant conductivity $\sigma_0$, and a secondary potential, $u_s$, resulting from the inhomogeneity such that $u = u_p + u_s$. The constant conductivity $\sigma_0$, is chosen as the local conductivity at the current source electrode as suggested by S. Zhao & M.J. Yedlin (1996). Based on this approach the governing equation (eq. 1) can be written as in eq. (5), wherein the singular delta source on the right side of the equation gets replaced as

$$\nabla \cdot (\sigma \nabla u_s) = \nabla \cdot \left((\sigma_0 - \sigma)\nabla u_p\right) \quad \text{in } \Omega. \quad (5)$$

The equivalent equation, considering a Fourier transform, for a domain with homogeneous conductivity in the $y$-direction, for wave number $k_y$, is given as

$$\widetilde{\nabla} \cdot \left(\sigma\widetilde{\nabla}\widetilde{u}_s\right) - k_y^2\sigma\widetilde{\nabla}\widetilde{u}_s = \widetilde{\nabla} \cdot \left((\sigma_0 - \sigma)\widetilde{\nabla}\widetilde{u}_p\right) - k_y^2(\sigma_0 - \sigma)\widetilde{\nabla}\widetilde{u}_p \quad \text{in } \Omega. \quad (6)$$

The main idea behind this split is that the singular part of the potential is captured by the primary potential (analytical solution) and the residual secondary potential is captured by solving the partial differential equation (PDE) in eq. (6). An analytical expression for the primary potential only exists for the case of flat topography, i.e. a flat half-space, and for a homogenous conductivity profile it is given as $u_p = \frac{I}{2\pi\sigma_0 r}$, or its 2.5-D equivalent

$$\tilde{u}_p(x, k_y, z) = \frac{I}{2\pi\sigma_0} K_0(k_y r_{xz}), \tag{7}$$

where $K_0$ is the zeroth order Bessel's function of second kind and $r_{xz} = \sqrt{(x - x_s)^2 + (z - z_s)^2}$ is the distance from the current source to an arbitrary point in the domain in $x - z$ plane.

On the top surface, S. Penz et al. (2013) suggested a zero total flux BC, yielding

$$\frac{\partial u_s}{\partial n} = -\frac{\partial u_p}{\partial n} = \frac{I}{2\pi\sigma_0} \frac{(\mathbf{x} - \mathbf{x}_s) \cdot \mathbf{n}}{r^3} = u_p \frac{\cos\theta}{r} \quad \text{on } \Gamma_s. \tag{8}$$

In a 2.5-D setting, this BC can be shown to be

$$\frac{\partial \tilde{u}_s}{\partial n} = -\frac{\partial \tilde{u}_p}{\partial n} = \alpha \tilde{u}_p \quad \text{on } \Gamma_s, \tag{9}$$

where $\alpha = \frac{k_y K_1(k_y r_{xz}) \cos\theta}{K_0(k_y r_{xz})}$, $K_1$ is the first order Bessel's function of second kind and $\theta$ is the angle between the radial vector from the source to a point on the top surface boundary and the outward normal to the boundary. The mixed boundary condition for the secondary potential on the truncated distant boundary is then given as $\frac{\partial u_s}{\partial n} + \frac{\cos\theta}{r} u_s = -\left(\frac{\partial u_p}{\partial n} + \frac{\cos\theta}{r} u_p\right) = 0$, which can be written in a 2.5-D setting as

$$\frac{\partial \tilde{u}_s}{\partial n} + \alpha \tilde{u}_s = -\left(\frac{\partial \tilde{u}_p}{\partial n} + \alpha \tilde{u}_p\right) = 0 \quad \text{on } \Gamma_\infty. \tag{10}$$

This approach enabled the application of the singularity removal technique for non-flat surfaces, while retaining the analytical solution for the case of flat topography to compute the primary potential. Here, the secondary potential ($u_s$) captures the effect of both the inhomogeneity in the conductivity distribution in the domain and the non-flat top surface boundary. The analysis of S. Penz et al. (2013) shows:

1. The Neumann boundary condition on a smooth surface (see eq. 8) is non-singular when the curvature of the points on the smooth surface at which the BC is evaluated is small.
2. For all the Gauss quadrature points in finite elements with linear shape functions connected to the source node, the scalar product in eq. (8) is always zero.

A closer look at eq. (8) reveals that a singularity arises in the secondary potential Neumann BC near the source, $\frac{\partial u_s}{\partial n} = u_p \frac{cos\theta}{r} \to \infty$ as $r \to 0$, when the term $cos\theta$ is finite, i.e. the normal at a point on the top surface is not orthogonal to the radial vector from the source to that point. To suppress this singular behaviour, S. Penz et al. (2013) propose that the topography near the source location exhibit very low curvature so that $cos\theta \approx 0$. The same arguments apply in the 2.5-D setting also, where low topography curvature at the source implies $\alpha \approx 0$. Under this condition, the Neumann boundary value is close to zero and remains well-behaved. Practically, this is achieved by artificially smoothing the terrain locally around the source, allowing the analytical primary potential, originally derived for a flat surface homogeneous half-space, to approximately satisfy the boundary condition $\frac{\partial u_p}{\partial n} = 0$ in the immediate neighbourhood of the electrode. While this smoothing mitigates the singularity, it may require alteration of the true topography near the source.

Thus, the Penz approach performs well when the surface is sufficiently smooth and the curvature at the source is small; conditions under which the flat surface analytical primary field remains a reasonable approximation (S. Penz et al. 2013; Z. Ren & J. Tang 2014). However, when the topography exhibits sharp corners or abrupt slope changes at the source, this assumption ceases to be applicable and the resulting primary potential deviates significantly from the flat surface solution, leading to large modelling errors. In the next section, we show how sharp corners are not just topographical features but arise naturally due to finite element discretization of smooth curves by linear elements.

*Influence of surface curvature at source and discretization*

Consider the case where the singular current is applied directly at a finite element node on the surface of a domain invariant in the *y*-direction. In DC resistivity forward modelling, linear finite elements are generally preferred over higher order elements because they provide better computational efficiency for large modelling domains, allow efficient local mesh refinement near source singularities, and reduce the cost of repeated forward evaluations during inversion (B. Zhou & S.A. Greenhalgh 2001; C. Rücker et al. 2006). However, when the actual surface topography is smooth and has a curvature at the source location, discretisation with linear shape functions approximates the geometry as a set of piecewise linear segments, thereby introducing artificial slope discontinuities between elements. For surfaces with gentle curvature, the magnitude of these discontinuities remains small, but when the source lies in a region of high curvature, the resulting geometric approximation produces substantial changes in slope. Consequently, the mapped surface exhibits sharp corners.

The arbitrary topographic surface illustrated in Fig. 1 serves as a representative example of a curved surface. This topography, when mapped using linear finite elements, will consist of nodes that lie on the true topography and elements that do not map the topography exactly. The discrepancy between the topography and the linear finite element mesh increases as the curvature of the surface increases. It is obvious that linear discretization leads to the formation of a V-shaped geometry at the source location formed by the two adjacent elements connecting the node at source. There is an abrupt change in slope between these two elements. However, this abrupt change is invisible numerically; the radial vector is perpendicular to the normal to element and the scalar product in the modified surface boundary condition, $(\mathbf{x} - \mathbf{x}_s).\mathbf{n} = \vec{\mathbf{r}}.\hat{\mathbf{n}} = 0$. This creates a condition where $cos\theta = 0$ and the BC in eq. (8) (resp. eq. 9) is locally always fixed to be 0 on the two linear elements connecting the source node. This is not the actual BC that would be applicable if the adjoining elements could trace the actual surface topology accurately. The vanishing secondary potential Neumann BC artefact, that arises due to discretization mismatch, purely due to numerical reasons, suggests that the analytical solution $u_p$ (resp. $\tilde{u}_p$) that must be considered, must be one that conforms to the $\frac{\partial u_p}{\partial n} = 0$ $\left(\text{resp. } \frac{\partial \tilde{u}_p}{\partial n} = 0\right)$ condition on the V-shaped geometry of the surface. The analytical solution $u_p = \frac{I}{2\pi\sigma_0 r}$ and its Fourier transform (see eq. 7) conform to this condition (see W.M. Telford, L.P. Geldart & R.E. Sheriff 1990), but for a flat surface half-space, and not the V-shaped surface geometry under consideration here. The modified analytical solution for a 2.5-D problem with a V-shaped surface is developed in the next section.

Furthermore, it is a worthwhile exercise to investigate the effect of the order of finite element shape functions on the discretization mismatch and the consequential error in the finite element potential due to application of the flat half space analytical solution shown in eq. (7). As shown in Fig. 1, higher order elements are expected to conform better to the surface topography and the corresponding potential errors are expected to be lower. These errors are expected to increase as the curvature of the surface topography increases or the radius of curvature R decreases.

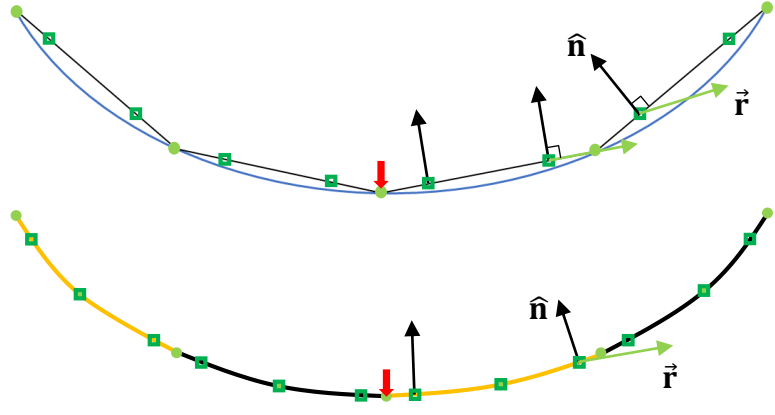

**Figure 1.** Top: curved surface traced by linear finite element (light green dots represent the nodes, green square - Gauss quadrature points, black/orange line – element, blue line – topographic surface). Bottom: curved surface traced by curved 2$^{nd}$ order element. The arrows denote the normal to the element and radial vector at corresponding Gauss point.

To assess the influence of topographic curvature, finite element simulations are performed on homogeneous domains (2000 m × 1000 m) with curved trench top surfaces generated using second and third order polynomial functions with various radii of curvature at the source location. The surface is initially discretized using finite elements with linear shape functions. The computation of normal to the surface (required by the BC in eq. 7) for the 1D boundary element is straightforward for the case of elements with straight edges. The normal is perpendicular to the line segment connecting the two end nodes and remains constant across the element (Fig. 1). For higher order elements with curved edges, the normal changes at each point along the curved element geometry (Fig. 1). Since the element geometry is defined by the shape functions, the tangent to the surface can be analytically obtained, and the normal is computed by rotating this tangent vector by 90°. The linear system of equations (eq. A6 in Appendix A) is solved to obtain the secondary potential, $\tilde{u}_s$. This result is combined with primary potential, $\tilde{u}_p$, which is then Fourier transformed back to the spatial domain to obtain the total potential in the domain.

Relative errors (per cent) in the electric potential at the top surface for topographies with varying curvature, generated using second and third order polynomial functions, are presented in Figs 2 and 3, respectively. As analytical solutions aren't available for the potential on a domain with these surface geometries, relative error is defined as the difference between the calculated total potential on these meshes (1st order Q4 elements and 2nd order Q9 elements with characteristic size of 0.25 m in the near-source region) and the calculated total potential on a very fine higher order mesh (3rd order Q16 elements with characteristic size of 0.125 m in the near-source region). On the very fine higher order mesh, the total potential ($\tilde{u}$) is obtained directly by numerically solving eq. (4) on the domain with a homogeneous conductivity, i.e. there is no primary-secondary split. Further details of the discretization, the BCs and the relative error computation are detailed in section 3.2. The errors in all cases show a constant trend. As expected, the computed electrical potential errors increase as the curvature at the source becomes more pronounced (Figs 2 & 3). With linear finite elements, the error is always greater than 1 per cent for all cases. When the same analysis is performed using second order elements whose geometry is defined by quadratic shape functions, a similar trend is observed, but the errors are lower across all cases (Figs 2 & 3). Moreover, when second order finite elements are used, the errors for a given radius of curvature are similar or slightly higher for topography generated using the second order polynomial function compared with that generated using the third order polynomial function. In contrast, when linear finite elements are employed, the errors become markedly larger for surfaces generated with the second order polynomial function (Figs 2 & 3). In conclusion, it is only in the limit, as the curvature at source decreases so that the surface near the source is approximately flat, that the flat half-space analytical solution is applicable. For general cases, especially when the surface is discretized with linear elements, we propose to use a modified analytical solution which incorporates the solid angle, $S$, subtended by the geometry at source, instead of the flat surface solid angle ($2\pi$), as discussed in the following section.

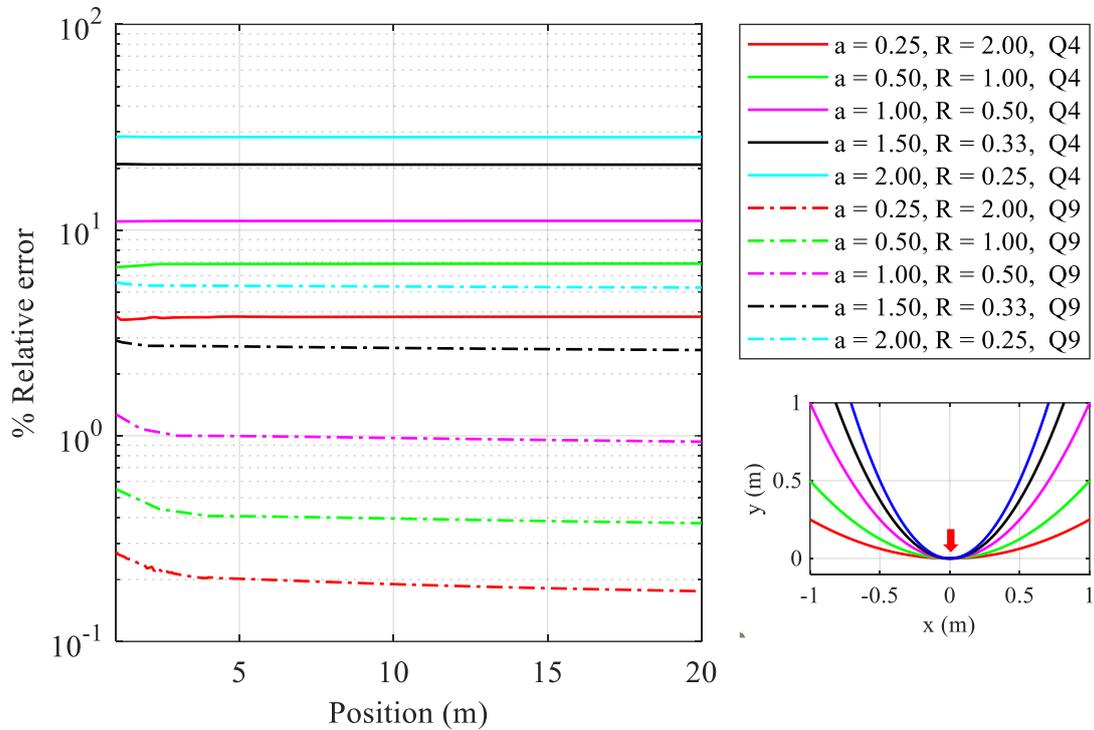

**Figure 2.** Variation in relative error with position along the top surface for different curvatures of the surface topography at the source location while applying the half-space primary potential and using first and second order elements. Surface topographies are generated using second order polynomial function ($y = ax^2$). Source is at (0, 0). R denotes the radius of curvature at source location.

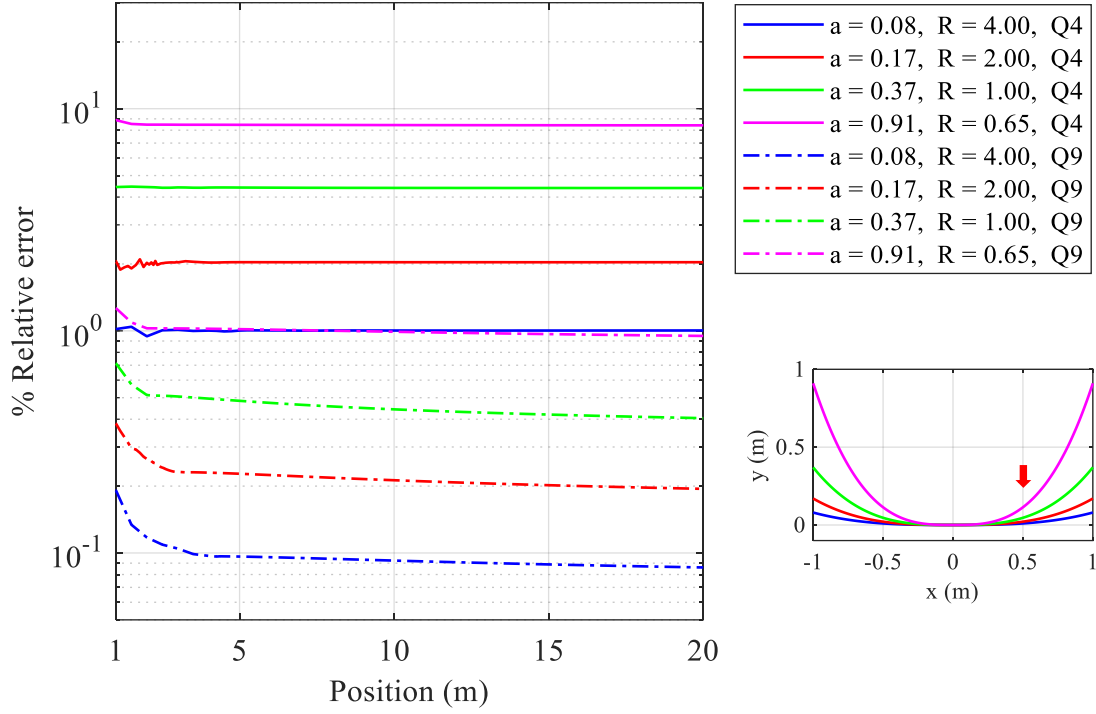

**Figure 3.** Variation in relative error with position along the top surface for different curvatures of the surface topography at the source location while applying the half-space primary potential and using first and second order elements. Surface topographies are generated using third order polynomial function (y = a|x|³). Source is at (0.5, 0). R denotes the radius of curvature at source location.

*2D: Potential due to a V-shaped wedge extending infinitely in the y-direction with a current source at its apex*

Working in spherical coordinates $(r, \theta, \varphi)$ with the polar axis taken along the $y$-axis, consider a spherical lune formed by two meridional planes of constant azimuthal angle $\varphi = \pm \frac{\beta}{2}$ respectively, as shown in Fig. 4. The domain of interest is the infinite region exterior to this spherical lune defined as $\Omega_{\text{exterior}} = \left\{ (r, \theta, \varphi) \middle| r > 0, 0 \leq \theta \leq \pi, \frac{\beta}{2} \leq \varphi < 2\pi - \frac{\beta}{2} \right\}$. The singular current is imposed at the origin. The meridional planes represent the earth's surface and act as insulating boundaries. As these planes pass through the origin, any point **x** on them satisfies the condition $\mathbf{x} \cdot \mathbf{n} = 0$, where **n** is the normal vector to these planes. This implies that the radial unit vector associated w.r.t **x**, i.e. $\hat{e}_r = \frac{\mathbf{x}}{r}$, where $r = \|\mathbf{x}\|_2$, is orthogonal to the normal to the plane because

$$\mathbf{n} \cdot \hat{e}_r = \frac{\mathbf{n} \cdot \mathbf{x}}{r} = 0. \tag{11}$$

It is easy to show that radial current density lines i.e. $\mathbf{J} = J_r(r)\hat{e}_r$, satisfy the boundary condition $\mathbf{J}.\mathbf{n} = 0$. This implies that a purely radial current density field satisfies the Neumann boundary conditions on the top-surface and we seek to find the radially symmetric current density field in $\Omega_{exterior}$. In the region away from the source with a homogeneous isotropic conductivity ($\sigma_0$), the current density follows Laplace's equation:

$$\nabla.\mathbf{J} = \frac{1}{r^2}\frac{d(r^2 J_r(r))}{dr} = 0. \tag{12}$$

Integrating this equation gives

$$r^2 J_r(r) = c_1, \tag{13}$$

where, $c_1$ is a constant which needs to be determined. The total current through this exterior spherical region is obtained from the integral over an area element $d\mathbf{A}$ at a radius $r$ from the source, given as

$$I = \iint_{\Omega_{exterior}} \mathbf{J}.d\mathbf{A} = \int_{\frac{\beta}{2}}^{2\pi - \frac{\beta}{2}} \int_0^{\pi} J_r(r)\hat{e}_r \cdot r^2 \sin\theta\, \hat{e}_r\, d\theta\, d\varphi = J_r(r)r^2 S, \tag{14}$$

where $S$ is the solid angle of the exterior spherical region which can easily be shown to be $S = 4\pi - 2\beta$. From eqs (13) and (14), the constant $c_1$ can be computed as

$$c_1 = r^2 J_r(r) = \frac{I}{S}, \tag{15}$$

and

$$J_r(r) = \frac{I}{Sr^2}. \tag{16}$$

Using Ohm's Law and the fact that for a purely radial current density field $\nabla u_p = \frac{du_p}{dr} \hat{e}_r$, we can write

$$-\sigma_0 \frac{du_p}{dr} = \frac{I}{Sr^2} \Rightarrow \frac{du_p}{dr} = -\frac{I}{\sigma_0 Sr^2}, \tag{17}$$

which on integrating w.r.t $r$ gives

$$u_p(r) = \frac{I}{\sigma_0 Sr} + c_2. \tag{18}$$

Imposing the boundary condition $u_p(r \to \infty) = 0$, implies $c_2 = 0$. This gives the final formula for the potential at any point in the exterior spherical region cut by two meridional planes as

$$u_p(\mathbf{x}) = u_p(r) = \frac{I}{\sigma_0 Sr}, \tag{19}$$

or equivalently in 2.5D as

$$\tilde{u}_p(x, k_y, z) = \frac{I}{\sigma_0 S} K_0(k_y r_{xz}). \tag{20}$$

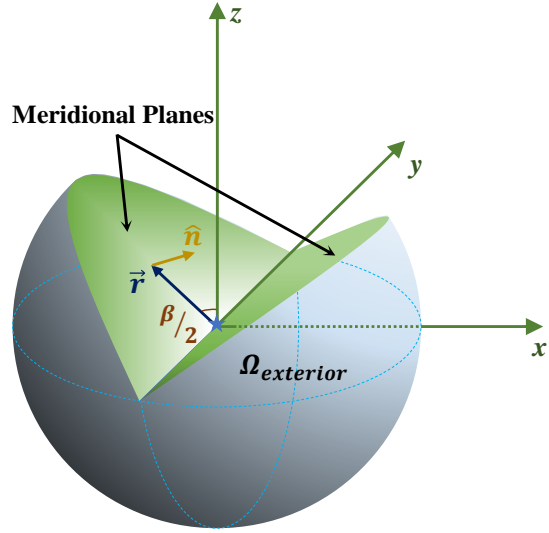

**Figure 4.** Sketch showing a spherical lune.

The analytical solution in eqs (19) and (20) take into account the volume of earth injected with current, through the solid angle $S$; for a flat or inclined half-space, $\beta = \pi$ and $S = 2\pi$, which returns the flat surface half-space analytical solution. Implementation of this "Modified half-space" analytical primary potential in place of the flat half-space solution naturally implies that the zero-flux Neumann BC, which is set-up as a numerical artefact in linear finite element discretization, is implicitly obeyed. This new primary potential $\tilde{u}_p$ (eq. 20) is used in eq. A6 (see appendix) to obtain the secondary potential $\tilde{u}_s$ which is then inverse Fourier transformed and added to $u_p$ (eq. 19) to obtain the total potential. This entire process including pre-processing, finite element assembly and linear system solution (eq. A6), using the UMFPACK library, is implemented in MATLAB.

## 3  NUMERICAL EXAMPLES

In order to verify the algorithm, the modelling error in the electric potential is first evaluated for two configurations with known analytical solutions: a flat homogeneous half space and a two layered subsurface. The proposed method is then examined for its accuracy under non-flat topographic conditions. Two representative surface geometries are considered: (i) a V-shaped trench and (ii) a sinusoidal hill-valley profile. All meshes for these examples are generated using the open source mesh generator Gmsh (C. Geuzaine & J.F. Remacle 2009) which has been used in several forward modelling codes (R. Schaa, L. Gross & J. Du Plessis 2016; T.C. Johnson, G.E. Hammond & X. Chen 2017; C. Rücker, T. Günther & F.M. Wagner 2017).

### 3.1 Verification with flat homogenous and 2-layered case

*Flat homogenous model*

The first verification model is a flat homogenous half-space of resistivity 10 Ωm. A unit point current source is placed at the origin (0, 0, 0), and surface potentials are computed along the top boundary. The computational domain is set to 2000 m × 1000 m to minimise the influence of boundary condition at the far edges. The near-source region, 30 m horizontally on either side and downward, is discretised using a uniform unstructured grid (characteristic element side length = 0.5 m), while a graded (characteristic element side length changes from 0.5 m to 62.5 m at the distant boundary) unstructured mesh is employed in the remaining domain. Quadrilateral finite elements of first, second, and third order (Q4, Q9, and Q16) having 4, 9, and 16 nodes respectively are used for the finite element analysis. Forward and inverse Fourier transforms use 17 wavenumbers, determined through a combination of 12-point Gauss-Legendre quadrature and 5-point Gauss-Laguerre quadrature following the approach detailed in A. Kemna (2000). These 17 wavenumbers are used in all subsequent examples. The relative error, computed against the analytical solution, is shown in Fig. 5. Errors remain below 0.1 per cent near the source and below 0.5 per cent elsewhere.

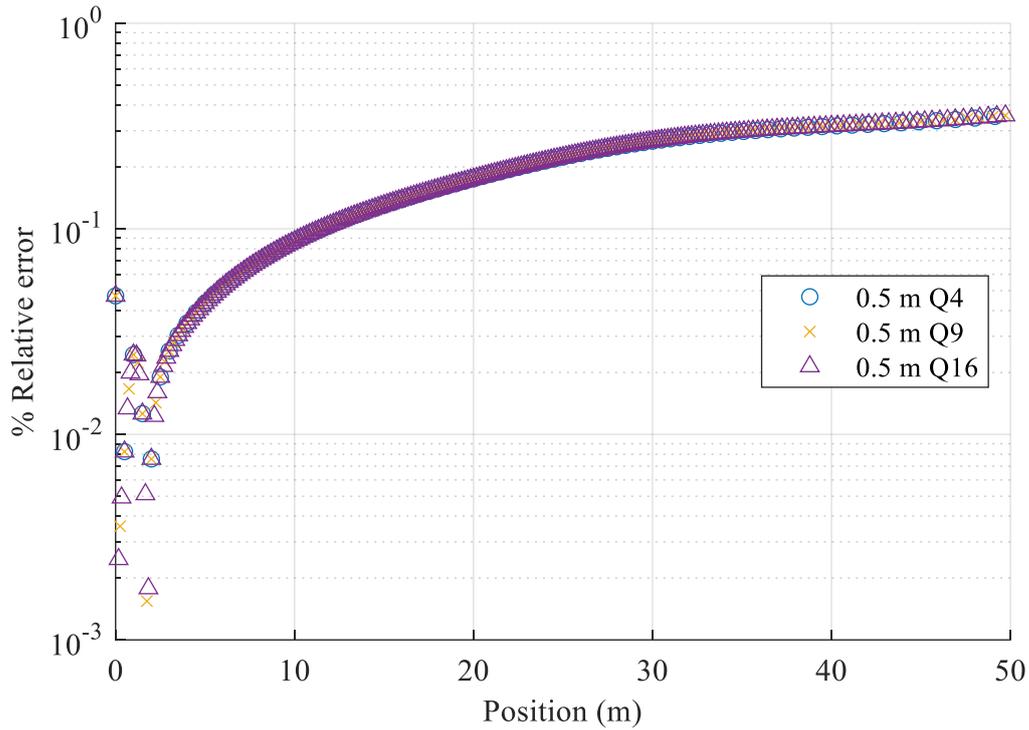

**Figure 5.** Per cent relative error for flat homogenous model. Lengths in the legend refer to the characteristic length of the side of an element in the uniform unstructured grid in the near-source region.

*Flat 2-layered model*

The second model considered for verification is a two layered model (Fig. 6). Surface potential is computed for a point source placed at the ground surface. The upper layer is 10 m thick with a conductivity of 1 Sm$^{-1}$ ($\rho$ = 1 $\Omega$m) and the lower layer has a conductivity 0.05 Sm$^{-1}$ ($\rho$ = 20 $\Omega$m). The computational domain and meshing strategy are identical to the previous case. Errors, calculated with respect to the analytical solution reported in W.M. Telford et al. (1990), remain well below 0.1 per cent (Fig. 7).

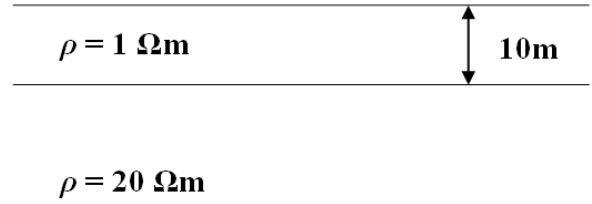

**Figure 6.** Flat 2-Layered model

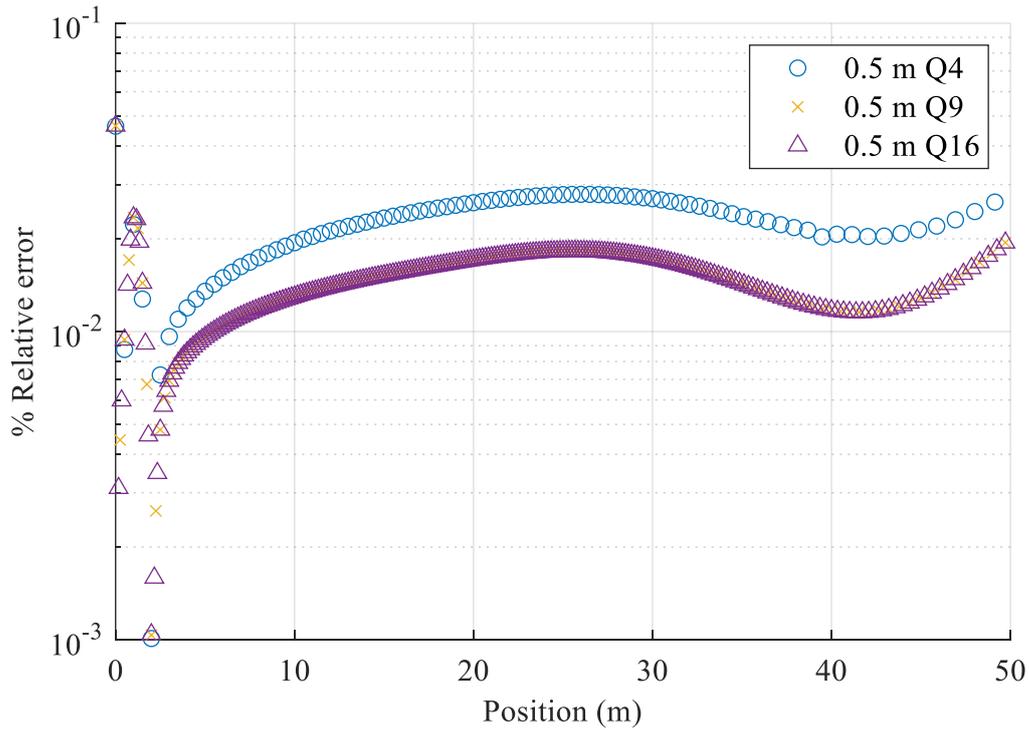

**Figure 7.** Per cent relative error for flat 2 Layered model. Lengths in the legend refer to the characteristic length of the side of an element in the uniform unstructured grid in the near-source region.

## 3.2 True solution for non-flat geometry

Since analytical solutions are not available for non-flat surfaces, the true solution is obtained following the approach of C. Rücker et al. (2006). In their method, the total potential of a homogeneous terrain was computed by solving the governing equation (eq. 4) using the finite element method with second order elements and a highly refined mesh around the source to accurately capture the singularity. The resulting numerically computed potential served as the primary field in their study. A similar strategy has been used by other researchers (Z. Ren & J. Tang 2014; Z. Ren et al. 2018a), and the relative error is computed with respect to the numerically derived reference solution. In the present work, this numerical approach is adopted to obtain the true potential for all undulating surface examples.

To identify the optimal mesh refinement and element order required to generate a sufficiently accurate true solution, several meshes with varying refinement levels in the near-source region and different element orders are examined for a flat homogeneous half-space. C. Rücker et al. (2006) employed linear and quadratic elements, applying *p*- and *h*-refinement strategies separately, and showed that *p*-refinement yielded substantially higher accuracy than *h*-refinement. Building on these insights, the present study implements a combined refinement strategy that uses both *p*- and *h*-refinement with quadratic and cubic element types. Fig. 8 illustrates a representative mesh with fine discretization extending 30 m around the source and a graded mesh outward. In this numerical study, the 60 m × 30 m near-source region is discretized using elements of characteristic side lengths of 0.125 m, 0.25 m and 0.50 m, and the element size increases from the characteristic side length at near-source region boundary to 62.5 m at the distant boundary in a gradual manner. The domain size is 2000 m × 1000 m. Table 1 summarizes the mesh configurations considered. Their accuracy is then evaluated by comparing the computed potentials with the analytical solution. Relative errors are evaluated up to 10 m from the source at 0.5 m intervals, and based on the results in Fig. 9, a mesh with 0.125 m element size near the source and third-order (Q16) elements is selected for generating the true potential for all non-flat models, as it provides low errors both near and away from the source.

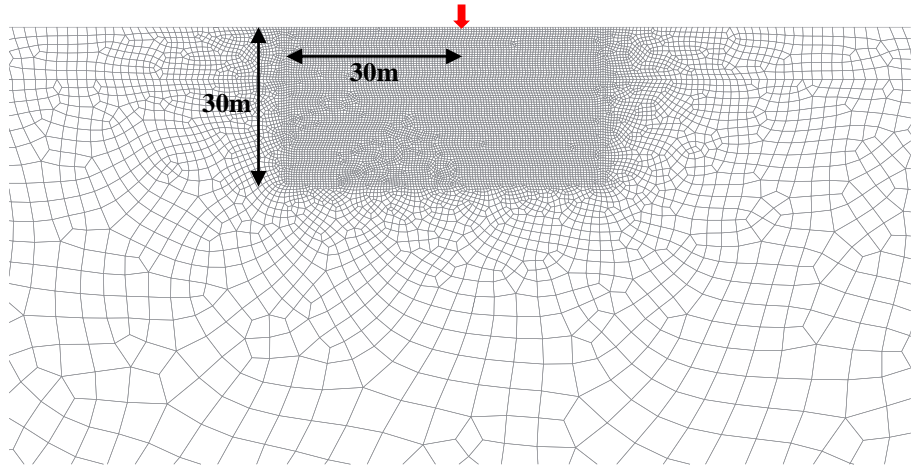

**Figure 8.** Section of mesh with a finely discretised near-source region. The red arrow denotes the location of the source electrode (0, 0). Characteristic element side length in the near-source region is 0.5 m.

**Table 1.** Details of various meshes generated

| S. No. | Mesh | Element size near source | Mesh order | No. of Elements/Nodes |
| --- | --- | --- | --- | --- |
| 1 | 0.50 m Q9 | 0.50 | 2 | 13,202/53,553 |
| 2 | 0.25 m Q9 | 0.25 | 2 | 49,085/197,409 |
| 3 | 0.125 m Q9 | 0.125 | 2 | 102,548/411,341 |
| 4 | 0.50 m Q16 | 0.50 | 3 | 13,202/119,935 |
| 5 | 0.25 m Q16 | 0.25 | 3 | 49,085/443,368 |
| 6 | 0.125 m Q16 | 0.125 | 3 | 102,548/924,655 |

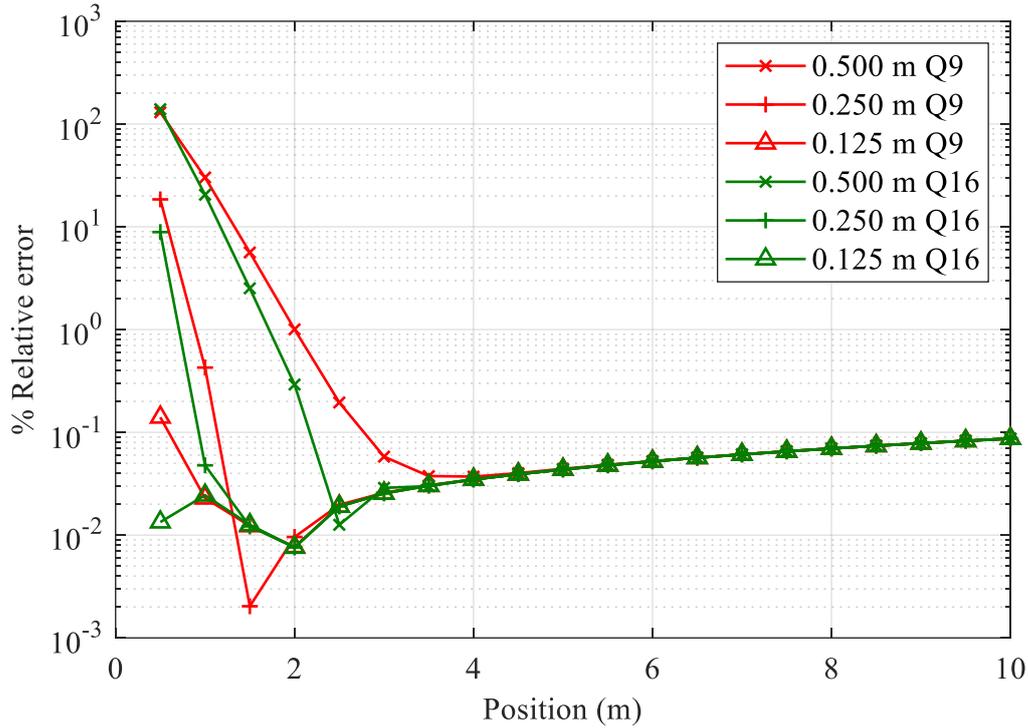

**Figure 9.** Relative error of numerical solution with respect to analytical solution in a flat surface half-space for different fineness of mesh near the source and order of finite elements. Lengths in the legend refer to the characteristic length of the side of an element in the uniform unstructured grid in the near-source region.

## 3.3 V-trench model

A 4 m deep V-shaped trench with side slopes inclined at 15° to the horizontal (Fig. 10) is selected as the first model for verification on non-flat topography. The trench slopes meet the flat surface 10 m from the vertex. A uniform structured mesh is used in the near-source region of the unit current source placed at the trench vertex, and both the element size in this refined near-source region and the global element order are varied to assess their influence on the computed potential. The true potential is obtained numerically using the total potential approach as explained in the previous section, employing a highly refined 0.125 m structured mesh near the source together with third order (16-node) quadrilateral elements. The solid angle subtended by the surface at the source is $S = 2.33\pi$, which is substituted into the analytical expression (eq. 19) to compute the primary potential. This modified primary potential is then used in the secondary potential formulation with the boundary conditions described earlier. The resulting total potential is compared against the true solution. Fig. 11 compares the numerical errors in potential from the proposed method (modified half-space primary potential) using various meshes, with those obtained using the approach proposed by S. Penz et al. (2013), wherein the primary potential corresponds to a flat half-space. The approach using the flat half-space solution exhibits significantly higher errors, since the topography at the source location exhibits a sharp slope variation, whereas the method relies on the analytical solution derived for flat surface. In contrast, the modified approach yields errors below 0.1 per cent. The error is observed to reduce with refinement of mesh in the vicinity of the source and with increasing order of mesh. Nevertheless, the error is satisfactorily lower even with a coarse mesh (near field element size of 1 m) with linear quadrilateral elements (Q4). The error plot exhibits small fluctuation near 10 m from the source, which can be attributed to the abrupt slope change at the location, but the magnitude, in all cases, remains below 0.1 per cent.

To assess the influence of the proximity of slope discontinuity to the source, three trenches with identical solid angle but different widths, 8 m, 20 m, and 40 m, are analysed, placing the slope change at 4 m, 10 m, and 20 m from the source, respectively. Each model is computed using the same framework with second order elements and a 0.5 m refined mesh in the near-source region. As shown in Fig. 12, all cases display minor fluctuations in error at the slope change location; however, the errors remain below 0.01 per cent across the domain. This indicates that the distance of the geometric discontinuity from the source has negligible impact on the overall accuracy of the computed potentials.

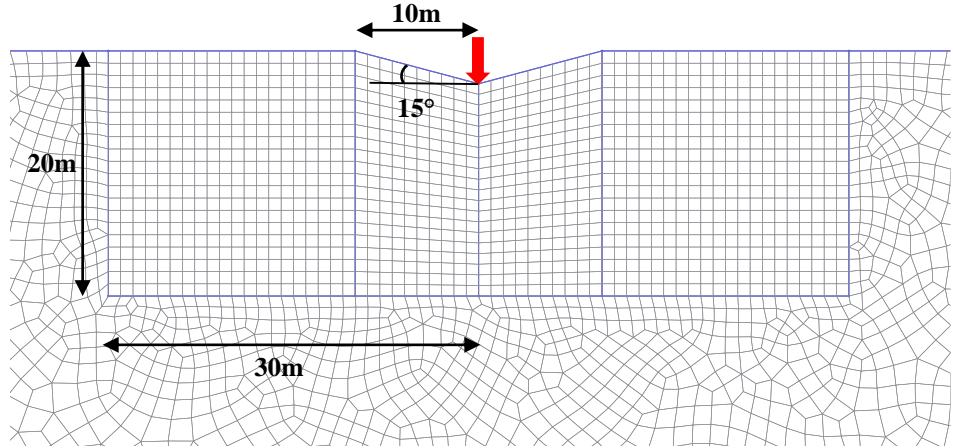

**Figure 10.** Finite element mesh of the V-trench model geometry with structured mesh in the near-source region and graded unstructured mesh extending out to the distant boundary.

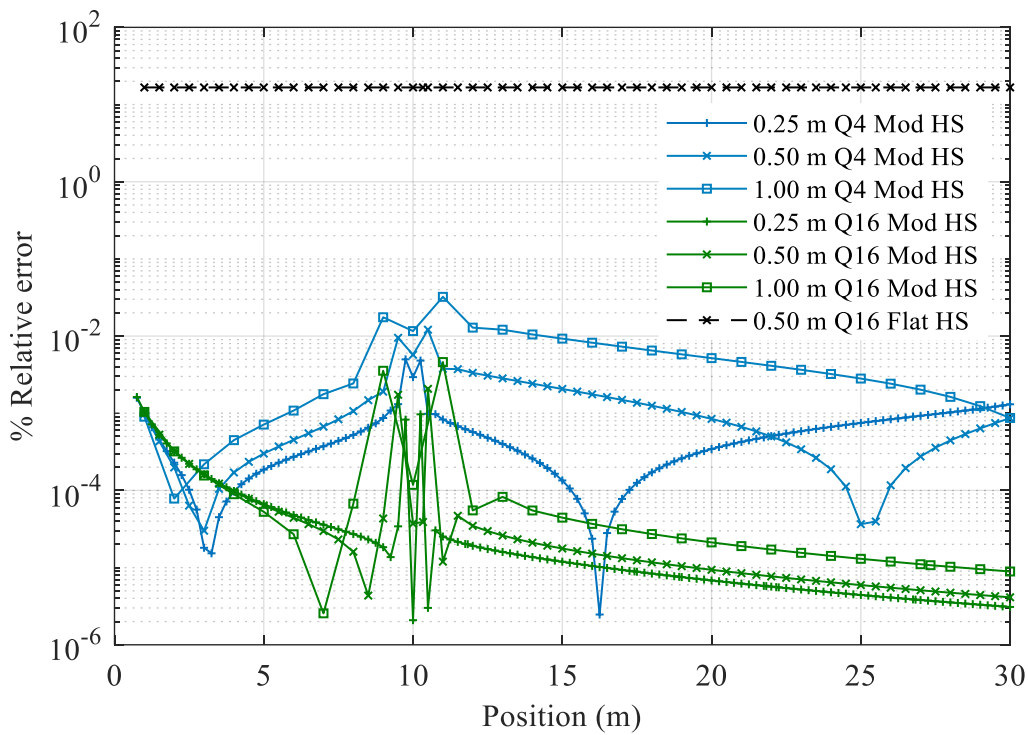

**Figure 11.** Relative error in total potential computed using the modified half-space primary potential and the half-space primary potential for a V-Trench topography. Lengths in the legend refer to the characteristic length of the side of an element in the uniform structured grid in the near-source region.

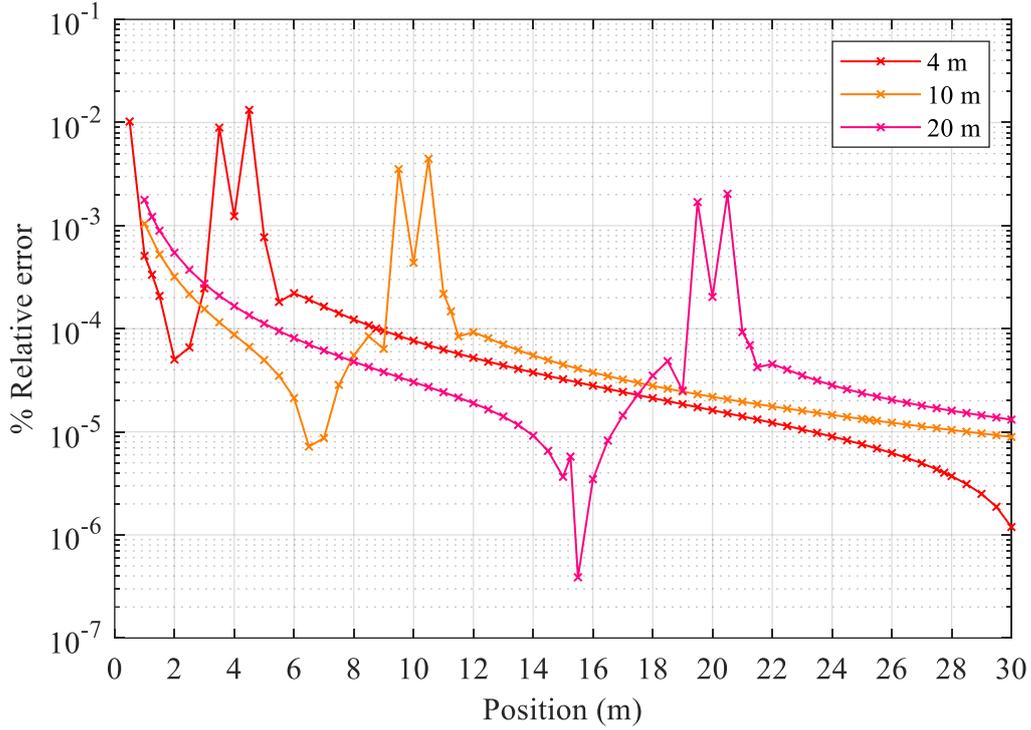

**Figure 12.** Variation in relative error for different trench widths computed using meshes with quadratic elements (Q9) and element size of 0.5 m in near-source region.

From Fig. 11 and Fig. 2, it is evident that the error remains constant when the flat half-space primary potential is used in cases in which the source is located at a slope discontinuity. To understand this, we compare the total potential obtained by solving the problem in eq. (6) with BC's in eqs (9) and (10). The linear system is written in terms of the total potential and the primary potential as shown in eq. (A7) (Appendix A), from where the linear relation $\tilde{u} = \mathbf{G}\tilde{u}_p$ can be obtained (see eq. A8 in Appendix A). Here $\tilde{u}_p$ denotes the modified half-space primary potential computed using the solid-angle-based formulation developed in eq. (20) in this study. Let $\tilde{u}_p^0$ be the flat half-space primary potential (eq. 7), which can be written in terms of the solid angle as,

$$\tilde{u}_p^0 = \frac{I K_0(k_y r_s)}{2\pi \sigma_0} = \frac{S}{2\pi} \tilde{u}_p. \tag{21}$$

The total potential, $\tilde{u}$, obtained with the modified-half space primary potential exhibits relative errors on the order of $10^{-3}$ and is therefore taken as an accurate approximation of the true total potential. The relative error in the solution ($\tilde{u}^0$), computed using the flat half-space primary potential can then be redefined with respect to $\tilde{u}$ as,

$$\left|\frac{\tilde{u}-\tilde{u}^0}{\tilde{u}}\right| = \left|\frac{\mathbf{G}\tilde{u}_p - \mathbf{G}\tilde{u}_p^0}{\mathbf{G}\tilde{u}_p}\right| = \left|\frac{\mathbf{G}\tilde{u}_p - \mathbf{G}\frac{S}{2\pi}\tilde{u}_p}{\mathbf{G}\tilde{u}_p}\right| = \left|1 - \frac{S}{2\pi}\right|. \tag{22}$$

The expression in eq. (22) shows that the error depends solely on the solid angle subtended by the surface topography at the current source location and is independent of the distance between the source and the potential electrode. Consequently, the error in the solution associated with the half-space primary potential remains constant with respect to the measurement position. For the V-trench example, eq. (22) yields errors of 16.5 per cent, which matches the per cent relative error observed in Fig. 11.

### 3.4 Sinusoidal hill-valley model

A sinusoidal hill-valley model (see Fig. 13) is selected as the second numerical example. A point source is placed in the valley at (-10, -4), and the potential is measured along the ground surface. The radius of curvature of the surface topography at the source location is 10.13, which indicates a small curvature. A uniform unstructured mesh is used in the near-source region, with a graded unstructured mesh elsewhere in the domain. The element size and the order of the elements in the uniform region are varied. For meshes composed of linear Q4 elements or straight-edged higher order elements, the discretization does not conform exactly to the sinusoidal surface. Consequently, the element edges subtend a solid angle at the source that differs from $2\pi$. In the proposed approach that uses the modified half-space primary potential, this solid angle is computed for each of the meshes and incorporated into the analytical expression for the modified half-space primary potential. In contrast, curved-edged higher order elements more accurately capture the geometry of the topographic surface, and for the present sinusoidal geometry, the solid angle at the source approaches the flat surface value ($S \approx 2\pi$). The true solution for this model is computed using the total potential approach with a very fine (0.125 m) mesh in the near-source region and third order (Q16) finite elements.

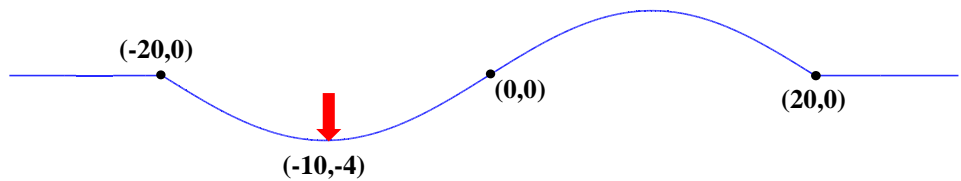

**Figure 13.** Sinusoidal hill-valley topography created using a sinusoidal function $[4\sin(\frac{2\pi x}{40})]$

The first set of analyses employs only straight-edged elements of all three orders. The errors associated with the flat half-space solution are relatively high (approximately 2 per cent), whereas the approach using the modified half-space primary potential yields low errors of approximately 0.1 per cent (Fig. 14). The higher errors in the prior case occur because the true primary potential deviates noticeably from that for a flat surface. With increasing mesh refinement near the source, the discretized geometry approximates the smooth sinusoidal surface more closely and becomes nearly flat at the source location, reducing the mismatch with the flat surface analytical solution. This leads to lower, but still significant errors of roughly 1 per cent. Although substantially finer meshes would further reduce these discrepancies and thus the error, they also greatly increase computational cost. In contrast, the proposed approach maintains low error levels even with coarse meshes (characteristic near-source element side length of 1 m) using linear elements, and the accuracy improves with higher order elements and finer discretization near the source (Fig. 15). Across all such cases, the proposed method produces errors consistently at or below 0.1 per cent.

Further analysis is performed using curved-edge higher order elements for the same topography and source configuration. As the element order and mesh refinement increase, the discretization conforms more accurately to the sinusoidal ground surface. Since the considered sinusoidal topographic surface is nearly flat at the source location, the primary potential computed from the flat-surface analytical expression ($S = 2\pi$) provides a close approximation to the true primary potential and consequently, the solutions considering the flat half-space and modified half-space primary potential converge. When curved-edged higher order elements are employed with characteristic element side length of 0.5 m in the near-source region, the errors remain at or below 0.1 per cent (see Fig. 16); coarser meshes in the near-source region (characteristic element side length of 1 m), produce higher error, reaching approximately 1 per cent, in the vicinity of the source. Interestingly, comparing the errors in Fig. 15 and Fig. 16 shows that simulations using higher-order curved-edge elements (with a characteristic element side length of 0.5 m), together with the flat half-space analytical solution, produce larger errors than simulations using higher-order straight-edged elements of a similar characteristic size, with the modified half-space solution. This reveals that the flat half-space assumption for the sinusoidal valley topography at $x = -10$ m is not entirely accurate.

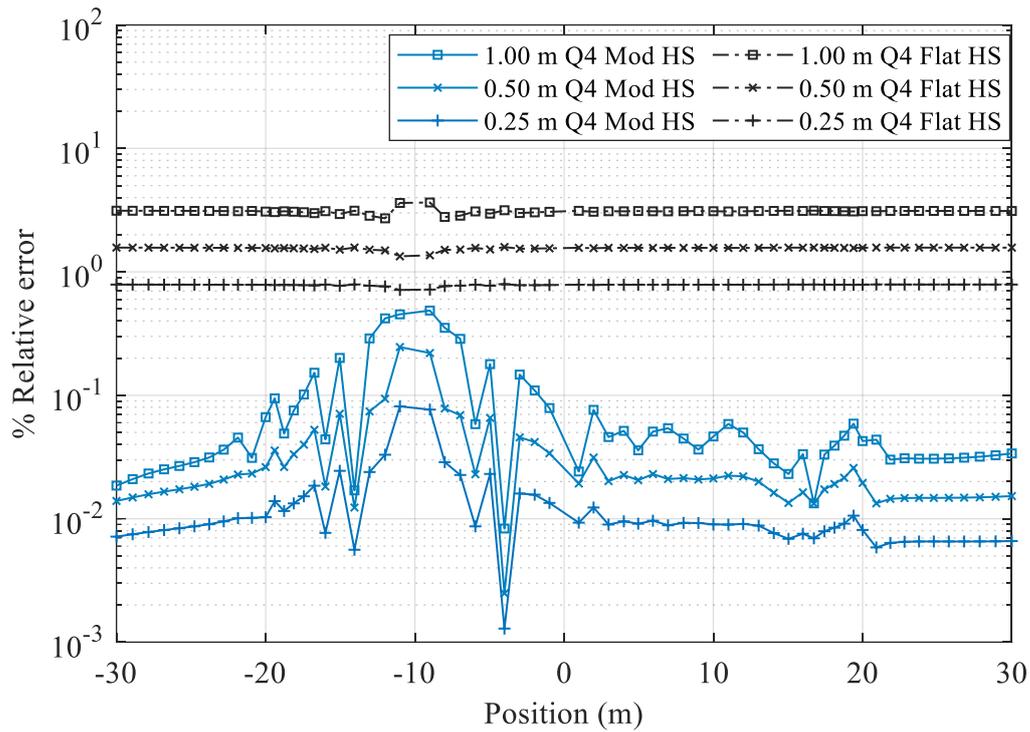

**Figure 14,** Relative error in total potential computed using the modified half-space primary potential and the half-space primary potential for a sinusoidal hill-valley topography discretized by linear elements. Source is at $x = -10$ m m. Lengths in the legend refer to the characteristic length of the side of an element in the uniform unstructured grid in the near-source region.

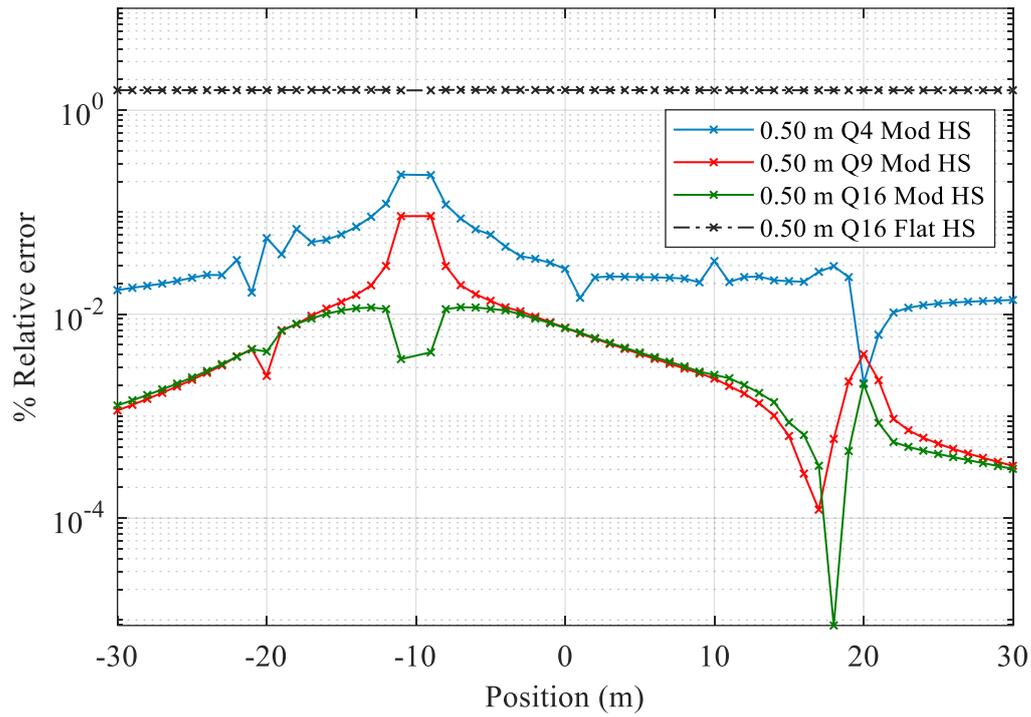

**Figure 15.** Relative error in total potential computed using the modified half-space primary potential and the flat half-space primary potential for a sinusoidal hill-valley topography with meshes of varying order and straight edged elements. Source is at $x = -10$ m. Lengths in the legend refer to the characteristic length of the side of an element in the uniform unstructured grid in the near-source region.

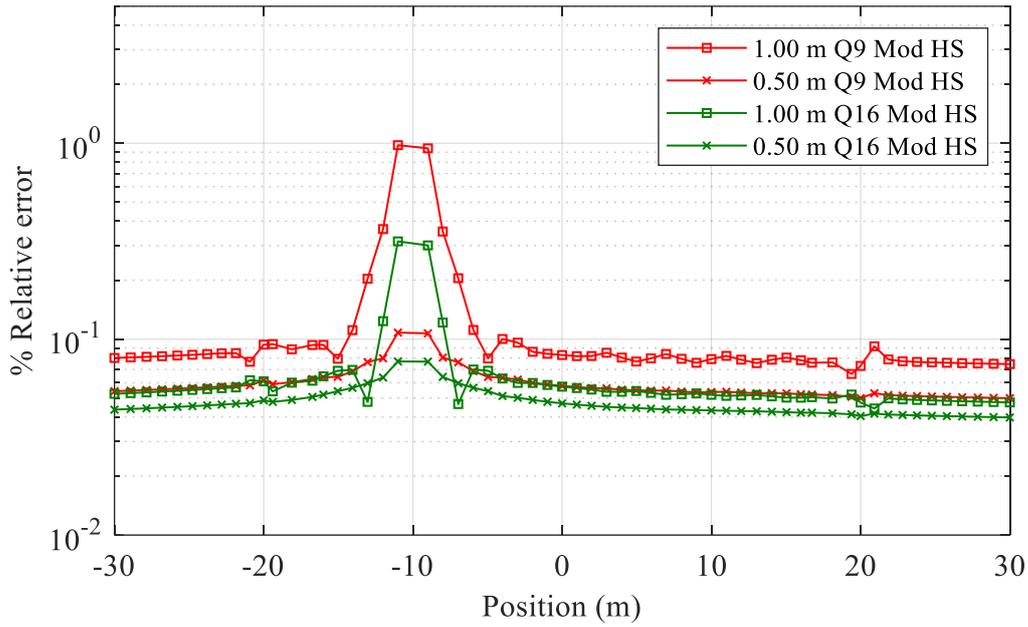

**Figure 16.** Relative error in total potential computed using the modified half-space primary potential and the flat half-space primary potential for a sinusoidal hill-valley topography with $S = 2\pi$ varying for higher order elements with curved edges. Source is at $x = -10$ m. Lengths in the legend refer to the characteristic length of the side of an element in the uniform unstructured grid in the near-source region.

Two additional scenarios are examined for the same sinusoidal topography but with different source positions. In the first scenario, the source is positioned at coordinates (0, 0), where the topographic surface is locally flat with no curvature or slope variation (radius of curvature, R = ∞). As a result, the solid angle equals $2\pi$ for both straight-edged and curved-edge element discretizations, since the surface (locally) exhibits no variation in slope. Consequently, both the approaches converge and yield identical results. Only the modified approach results are presented in Fig. 17, which show errors to be below 0.1 per cent for linear quadrilateral elements. In the second scenario, the source is placed at (-20, 0), at the junction between the sinusoidal surface and the adjacent flat ground. At this location, the topography exhibits an abrupt change in slope and thus even with mesh discretization that perfectly conforms to the surface, the solid angle deviates from 2π. Under such conditions, the solid angle must be explicitly calculated and incorporated into the primary potential computation, regardless of whether straight-edged or curved-edge higher order elements are used. This requirement is clearly reflected in the high errors produced while using flat half-space analytical solution (Fig. 18 and Fig. 19), which reach approximately 10 per cent. In the proposed approach, the solid angle is calculated for each mesh and used to compute the modified half-space solution, with which the error significantly reduces.

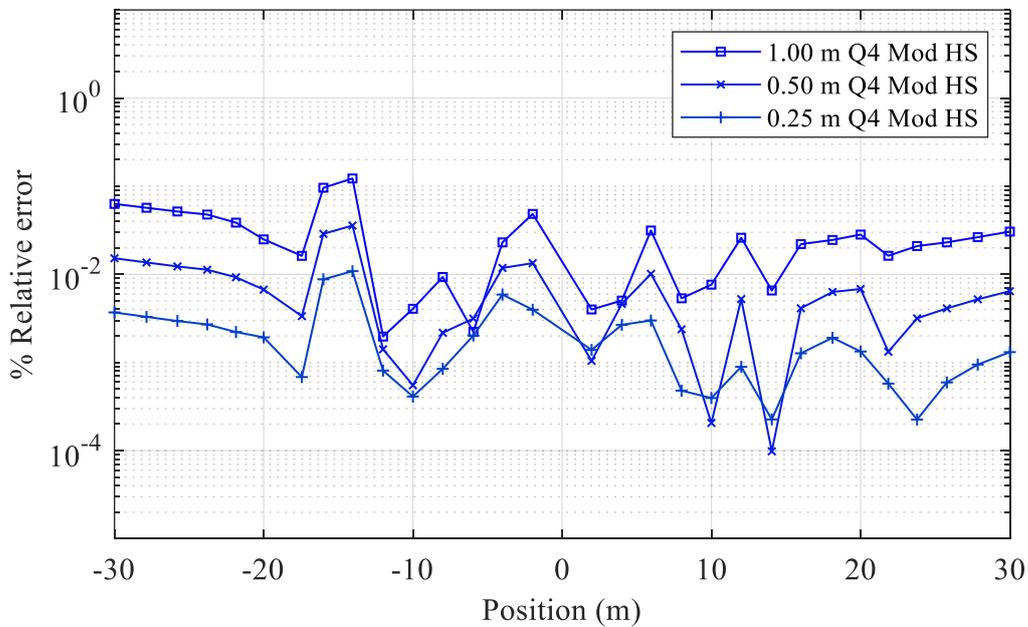

**Figure 17.** Relative error in total potential computed using the modified half-space primary potential for a sinusoidal hill-valley topography with source at (0,0) and solid angle $S = 2\pi$. Lengths in the legend refer to the characteristic length of the side of an element in the uniform unstructured grid in the near-source region.

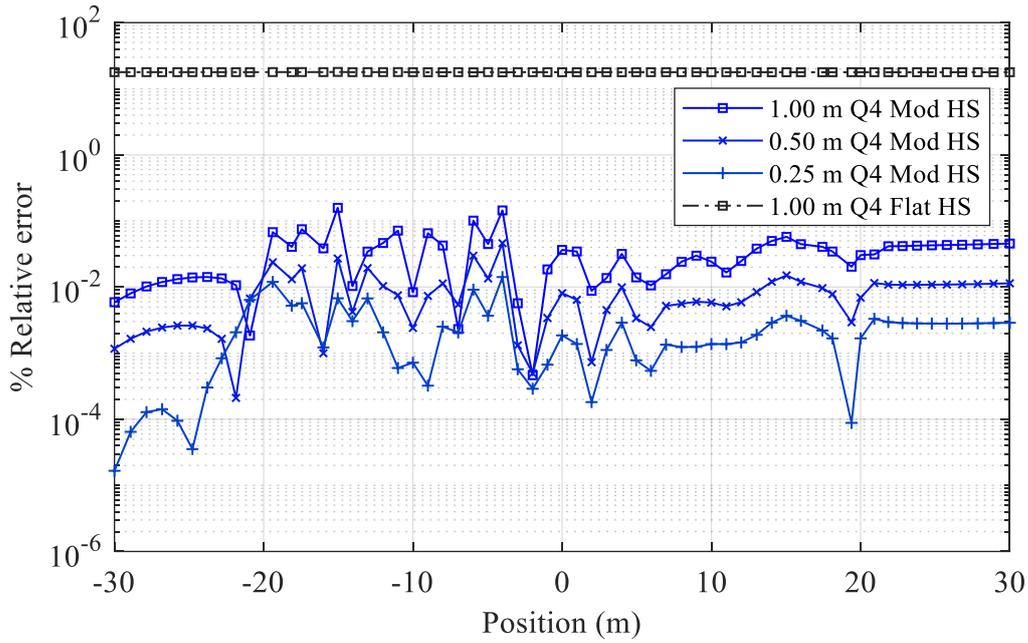

**Figure 18.** Relative error in total potential computed using the modified half-space primary potential and half-space primary potential for a sinusoidal hill-valley topography with source at (-20,0). Lengths in the legend refer to the characteristic length of the side of an element in the uniform unstructured grid in the near-source region.

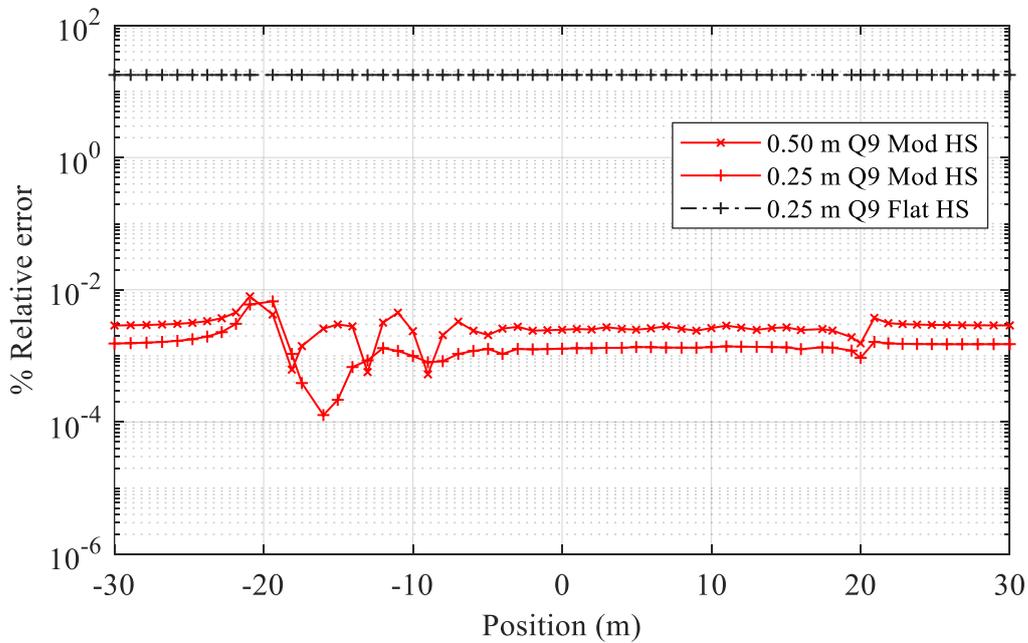

**Figure 19.** Relative error in total potential computed using the modified half-space primary potential and half-space primary potential for a sinusoidal hill-valley topography with source

at (-20,0), with a Q9 mesh having curved edges. Lengths in the legend refer to the characteristic length of the side of an element in the uniform unstructured grid in the near-source region.

## 4   SUMMARY AND CONCLUSIONS

A modified singularity removal technique for undulating topography has been presented. The study first examined the limitation of existing methods, demonstrating that surface curvature near the current source fundamentally governs modelling accuracy. The analysis confirmed that the secondary potential approach of S. Penz et al. (2013), although computationally efficient, is highly sensitive to local geometric conditions at the source location. When the ground surface is smooth or nearly planar, the analytical primary potential derived for a flat half space provides a reasonable approximation. However, when the actual topography is non-smooth and contains features such as sharp corners and slope discontinuities or when a smooth topography includes regions of high curvature that are discretized using linear finite elements, this assumption becomes invalid. The resulting mismatch produces significant modelling errors that persist even under mesh refinement. These findings confirm that the primary source of error stems from the mismatch between the analytical primary potential and the true geometric singularity imposed by the topography.

To address this limitation, a new analytical formulation for the primary potential is developed based on the solid angle subtended by the local topography at the source. This formulation ensures that the background model satisfies the boundary value problem locally at the singularity. A central contribution of this work is the derivation of an analytical expression for the potential generated by a point source located at the apex of a V-shaped wedge, which serves as a fundamental building block for extending the primary field representation to arbitrary surface geometries. Notably, the proposed approach circumvents the need to compute the singular primary potential using computationally intensive numerical schemes such as those employed by C. Rücker et al. (2006), M. Blome et al. (2009) and Z. Ren et al. (2018a). The method also eliminates the dependence on surface smoothness and avoids the necessity for artificial topography modification or extremely fine mesh refinement, thereby offering substantial improvements in computational efficiency without compromising accuracy. Moreover, the proposed method uses the exact solid angle subtended by linear and higher-order straight-edged mesh elements and makes no local-flatness assumptions about the topography, thereby allowing straight-edged (including higher-order) elements to produce more accurate solutions than curved-edge higher-order elements.

The numerical validation studies demonstrate that the solid angle based modified secondary potential formulation consistently yields very low modelling errors ($\leq 0.1$ per cent) across a variety of topographic configurations, including V-shaped trench and sinusoidal hill-valley surface, and for multiple source positions. Notably, the method performs well even with coarse linear finite element discretization. The findings highlight the method's practicality for 2.5-D modelling, where linear finite elements remain widely used due to their scalability for large domains and the computational demands of repeated forward evaluations during inversion.

In conclusion, the solid angle based analytical primary potential provides a robust and computationally efficient alternative to existing singularity removal strategies for complex topography. Its ability to maintain high accuracy across sharp geometric features and varying surface curvature, even with coarse linear discretization, makes it a strong candidate for integration into large-scale 2.5-D inversion frameworks. Future work may extend the approach to full 3-D forward modelling and examine how the primary potential for arbitrary 3-D surfaces can be formulated using the solid angle concept.

**CRediT AUTHORSHIP CONTRIBUTION STATEMENT**

**Naveen K**: Conceptualization, Data curation, Formal analysis, Investigation, Methodology, Software, Validation, Visualization, Writing - original draft; **Michael C Koch**: Conceptualization, Methodology, Software, Writing - review and editing; **Kazunori Fujisawa**: Conceptualization, Supervision, Writing - review and editing; **Arindam Dey**: Funding acquisition, Project administration, Resources, Supervision, Writing - review and editing; **Sreedeep S**: Project administration, Resources, Supervision, Writing - review and editing

**DECLARATION OF COMPETING INTEREST**

The authors declare that they have no known competing financial interests or personal relationships that could have appeared to influence the work reported in this paper.

**REFERENCES**

Blome, M., Maurer, H. & Kersten, S., 2009. Advances on 3D geoelectric forward solver techniques, Geophys. J. Int., 176, 740-752.

Coggon, J.H., 1971. Electromagnetic and electrical modeling by the finite element method, Geophysics, 36(1), 132-155.

Demirci, I., Erdogan, E. & Candansayar, M.E., 2012. Two-dimensional inversion of direct current resistivity data incorporating topography by using finite difference techniques with triangle cells: investigation of Kera fault zone in western Crete, Geophysics, 77(1), E67-E75.

Dey, A. & Morrisson, H.F., 1979. Resistivity modeling for arbitrarily shaped three-dimensional structures, Geophysics, 44(4), 753-780.

Dieter, K., Paterson, N.R. & Grant, F.S., 1969. IP and resistivity type curves for three-dimensional bodies, Geophysics, 34, 615-632.

Fox, R.C., Hohmann, G.W., Killpack, T.J., Rijo, L., 1980. Topographic effects in resistivity and induced-polarization surveys, Geophysics, 45, 75-93.

Geuzaine, C. & Remacle, J.F., 2009. Gmsh: a 3-D finite element mesh generator with built-in pre- and post-processing facilities, Int. J. Numer. Methods Eng., 79(11), 1309-1331.

Günther, T., Rücker, C. & Spitzer, K., 2006. Three-dimensional modelling and inversion of dc resistivity data incorporating topography – II. Inversion, Geophys. J. Int., 166(2), 506-517.

Holcombe, H.T. & Jiracek, G.R., 1984. Three-dimensional terrain corrections in resistivity surveys, Geophysics, 49, 439–452.

Johnson, T.C., Hammond, G.E. & Chen, X., 2017. PFLOTRAN-E4D: A parallel open source PFLOTRAN module for simulating time-lapse electrical resistivity data, Comput. Geosci., 99, 72-80.

Kemna A., 2000. *Tomographic inversion of complex resistivity: theory and application*, PhD thesis, Ruhr Universität Bochum, Bochum, Germany.

Loke, M.H., Chambers, J.E., Rücker, D.F., Kuras, O. & Wilkinson, P.B., 2013. Recent developments in the direct-current geoelectrical imaging method, J. Appl. Geophys., 95, 135-56.

Lowry, T., Allen, M.B. & Shive, P.N., 1989. Singularity removal: a refinement of resistivity modeling techniques, Geophysics, 54(6), 766-774.

Mufti, I.R., 1976. Finite-difference resistivity modeling for arbitrarily shaped two-dimensional structures, Geophysics, 41(1), 62-78.

Okabe, M., 1981. Boundary element method formulations theory arbitrary inhomogeneities problem in electrical prospecting, Geophys. Prospect., 29(1), 39-59.

Pan K.J. and Tang J.T., 2014. 2.5 D and 3 D resistivity modelling using an extrapolation cascadic multigrid method. Geophys. J. Int., 197, 1459-1470.

Penz, S., Chauris, H., Donno, D. & Mehl, C., 2013. Resistivity modelling with topography, Geophys. J. Int., 194(3), 1486-1497.

Pridmore, D.F., Hohmann, G.W., Ward, S.H. & Sill, W.R., 1981. An investigation of finite-element modeling for electrical and electromagnetic data in three dimensions, Geophysics, 46(7), 1009-1024.

Qiu, L.W., Liu, Z.G., Yao, H.B. & Tang, J.T., 2024a. DC3DPAFEM: an efficient and accurate 3D direct current resistivity anisotropic forward modeling software for complex geological settings, Comput. Geosci., 189, 105623.

Qiu, L.W., Tang, J.T. & Liu, Z.G., 2024b. An improved goal-oriented adaptive finite-element method for 3D direct current resistivity anisotropic forward modeling using nested tetrahedra, J. Appl. Geophys., 231, 105555.

Ren, Z. & Tang, J., 2014. A goal-oriented adaptive finite-element approach for multi-electrode resistivity system, Geophys. J. Int., 199(1), 136-145.

Ren, Z., Chen, H. & Tang, J., 2018a. Accurate volume integral solutions of direct current resistivity potentials for inhomogeneous conductivities in half space, J. Appl. Geophys., 151, 40-46.

Ren, Z., Qiu, L., Tang, J., Wu, X., Xiao, X. & Zhou, Z., 2018b. 3-D direct current resistivity anisotropic modelling by goal-oriented adaptive finite element methods, Geophys. J. Int., 212, 76-87.

Rücker, C., Günther, T. & Spitzer, K., 2006. Three-dimensional modelling and inversion of DC resistivity data incorporating topography – I. Modelling, Geophys. J. Int., 166, 495-505.

Rücker, C., Günther, T. & Wagner, F.M., 2017. pyGIMLi: An open-source library for modelling and inversion in geophysics, Comput. Geosci., 109, 106-123.

Schaa, R., Gross, L. & Du Plessis, J., 2016. PDE-based geophysical modelling using finite elements: examples from 3D resistivity and 2D magnetotellurics, J. Geophys. Eng., 13 (2), S59–S73.

Spitzer, K., 1995. A 3-D finite-difference algorithm for DC resistivity modelling using conjugate gradient methods, Geophys. J. Int., 123(3), 903–914.

Telford, W.M., Geldart, L.P. & Sheriff, R.E., 1990. Applied geophysics, Cambridge University Press.

Tsourlos, P.I., Szymanski J.E. & Tsokas, G.N., 1999. The effect of terrain topography on commonly used resistivity arrays, Geophysics, 64, 1357-1363.

Zhao, S. & Yedlin, M.J., 1996. Some refinements on the finite-difference method for 3-d dc resistivity modeling, Geophysics, 61(5), 1301-1307.

Zhou, B. & Greenhalgh, S.A., 2001. Finite element three-dimensional direct current resistivity modelling: accuracy and efficiency considerations, Geophys. J. Int., 145, 679-688.

APPENDIX A: *Finite Element Implementation*

Finite element method is used to compute the unknown secondary potentials in the domain generated due to the input point current source and with the boundary conditions at the surface and distant boundaries. Following the method of weighted residuals, eq. (6), the Fourier transformed form of eq. (5), is multiplied by a weighting function, $\omega$, on both the sides and integrated over the domain.

$$\int_\Omega \omega \left( \nabla \cdot (\sigma \nabla \tilde{u}_s) - k_y^2 \sigma \tilde{u}_s \right) d\Omega$$

$$= \int_\Omega \omega \left( \nabla \cdot \left( (\sigma_0 - \sigma) \nabla \tilde{u}_p \right) - k_y^2 (\sigma_0 - \sigma) \tilde{u}_p \right) d\Omega. \tag{A1}$$

The differential equation is thus converted to an integral form which has to be solved to obtain the secondary potential. Using Green's theorem to integrate by parts and rearranging the terms, we get,

$$\int_\Omega \left( (\nabla \omega \sigma \nabla \tilde{u}_s) + \omega k_y^2 \sigma \tilde{u}_s \right) d\Omega + \int_{\Gamma_\infty} (\omega \sigma \alpha \tilde{u}_s) d\Gamma \tag{A2}$$

$$= \int_\Omega \left( (\nabla \omega \sigma_0 \nabla \tilde{u}_p) + \omega k_y^2 \sigma_0 \tilde{u}_p \right) d\Omega + \int_{\Gamma_\infty} (\omega \sigma_0 \alpha \tilde{u}_p) d\Gamma$$

$$+ \int_{\Gamma_s} (\omega \sigma_0 \alpha \tilde{u}_p) d\Gamma - \int_\Omega \left( (\nabla \omega \sigma \nabla \tilde{u}_p) + \omega k_y^2 \sigma \tilde{u}_p \right) d\Omega$$

$$- \int_{\Gamma_\infty} (\omega \sigma \alpha \tilde{u}_p) d\Gamma.$$

By following Galerkin's approach, same function, $N$, is considered as weighting function and shape function. Further, discretising the domain into finite elements, the Fourier potential, $\tilde{u}$, can be approximated by expressing it as a combination of shape functions as below,

$$\tilde{u} = \sum_{j=1}^{n_{ne}} \tilde{u}_j N_j^e, \tag{A3}$$

where $\tilde{u}_j$ is the Fourier potential at $j^{th}$ node of an element, $N_j^e$ is the corresponding shape function and $n_{n_e}$ is the number of nodes in an element. Substituting for $\tilde{u}_s$, $\tilde{u}_p$ and $\omega = N$ into eq. (A2), and summing over all the elements, we get

$$\sum_e \left\{ \sum_{i,j} \left[ \int_{\Omega_e} \left( (\nabla N_i^{eT} \sigma \nabla N_j^e) + N_i^{eT} k_y^2 \sigma N_j^e \right) d\Omega + \int_{\Gamma_{e\infty}} (N_i^{eT} \sigma \alpha_\infty N_j^e) d\Gamma \right] \tilde{u}_s \right\}$$

$$= \sum_e \left\{ \sum_i \left[ \int_{\Omega_e} \left( (\nabla N_i^{eT} \sigma_0 \nabla N_j^e) + N_i^{eT} k_y^2 \sigma_0 N_j^e \right) d\Omega \right. \right.$$

$$+ \int_{\Gamma_{e\infty}} (N_i^{eT} \sigma_0 \alpha_\infty N_j^e) d\Gamma + \int_{\Gamma_{e_s}} (N_i^{eT} \sigma_0 \alpha_s N_j^e) d\Gamma$$

$$- \int_{\Omega_e} \left( (\nabla N_i^{eT} \sigma \nabla N_j^e) + N_i^{eT} k_y^2 \sigma N_j^e \right) d\Omega$$

$$\left. \left. - \int_{\Gamma_{e\infty}} (N_i^{eT} \sigma \alpha_\infty N_j^e) d\Gamma \right] \tilde{u}_p \right\} \tag{A4}$$

where $i, j = 1,\ldots$ up to the number of nodes in respective domain and boundary elements. The final matrix form after assembly of element level terms, is

$$[\mathbf{A}_{\sigma_d} + \mathbf{A}_{\sigma_\infty}]\tilde{u}_s = [\sigma_0(\mathbf{A}_{1_d} + \mathbf{A}_{1_\infty} + \mathbf{A}_{1_s}) - (\mathbf{A}_{\sigma_d} + \mathbf{A}_{\sigma_\infty})]\tilde{u}_p \tag{A5}$$

The domain and boundary matrices are summed and can be written in a concise form as below,

$$\mathbf{A}_\sigma \tilde{u}_s = [\sigma_0 \mathbf{A}_1 - \mathbf{A}_\sigma]\tilde{u}_p = \mathbf{b}. \tag{A6}$$

Matrix $\mathbf{A}_1$ is the stiffness matrix for homogenous ground with unit conductivity which has to be created only once. The stiffness matrix, $\mathbf{A}_\sigma$, is positive definite, symmetric and sparse, and the right-side term, **b**, is a column vector. The stiffness matrix is stored as a sparse matrix for efficient memory utilisation. The forward modeling algorithm is coded in MATLAB and eq.

(A6) is solved using the backslash operator, which automatically detects the type of matrix and selects optimal direct solver.

The linear system in eq. A5 can be rewritten in terms of the total potential as

$$[\mathbf{A}_{\sigma_d} + \mathbf{A}_{\sigma_\infty}]\tilde{u} = [\sigma_0(\mathbf{A}_{1_d} + \mathbf{A}_{1_\infty} + \mathbf{A}_{1_s})]\tilde{u}_p, \tag{A7}$$

which gives the linear relation between the total potential and the primary potential

$$\tilde{u} = [\mathbf{A}_{\sigma_d} + \mathbf{A}_{\sigma_\infty}]^{-1}[\sigma_0(\mathbf{A}_{1_d} + \mathbf{A}_{1_\infty} + \mathbf{A}_{1_s})]\tilde{u}_p = \mathbf{G}\tilde{u}_p. \tag{A8}$$

This formulation helps quantify and compare the errors in the total potential solution due to different choices of the primary potential $\tilde{u}_p$.